# Methods of Targets' Characterization

*Christelle* Stodel[1,*]

[1]Grand Accélérateur National d'Ions Lourds (GANIL), CNRS/CEA, Bvd Henri Becquerel, BP 55027, 14076 Caen CEDEX 5, France

**Abstract.** The paper reviews the characterization's techniques for solid targets used in nuclear physics with special emphasis on actinide targets. The determination of the thickness, isotopic and chemical composition are described for actinide targets during their fabrication process. Accent is given on their monitoring during irradiation.

## 1 Introduction

In nuclear physics, targets are an essential ingredient for the success of experiments. Precise knowledge of target properties like e.g., the thickness, its distribution (or homogeneity) over the surface and its purity either isotopically or chemically is irrevocable for proper analysis of experimental data.

Indeed, knowledge of the kind and level of target impurities is required for evaluating side reactions competitive to the one under study and in estimating the background level. In addition, the characterization of actinide targets is important as the base material consists often of various radioactive isotopes, including decay products. Although some nuclides are present in a very low amount, they can induce high activity due to their short half-life but also they might interfere the measurements. For instance, in the thermal-neutron induced fission reaction of $^{238}$U target cited in [1], the main contributions (at 37.3% and 5.3 % respectively) in the measured cross-section comes from the high fissility of $^{235}$U present at only 12 ppm in the raw material and from $^{239}$Pu formed after neutron capture in $^{238}$U.

The areal density (thickness) of targets and stripper foils is also a crucial parameter for nuclear physics experiments. As a matter of fact, it is taken into account in the kinematical characteristics of the produced nucleus of interest, which will be used to tune correctly the spectrometer in order to optimize its transmission. Moreover, the thickness of the targets influence directly the results of cross-sections measurements which are often required with uncertainties in the level of percent. P. Schillebeeckx et al. [2] and D. Sapundjiev et al. [3] report that the improvement in the quality of cross-section data in various fields (astrophysics, neutron-induced reaction for nuclear power production, fuel cycle) relies mainly in the high quality targets with known properties at any time. Another aspect to be considered is the variation of target properties during particle irradiation due to physical and chemical instability of the material, then the monitoring of various target performance parameters all along the experiments is essential.

This paper does not pretend to be exhaustive and read as a textbook, but recalls some general principles and results of the used techniques for which details are provided in the quoted bibliography. As mentioned, only active material and monitoring under irradiation are under consideration, additional information on target characterization can be found in [4-6] and references therein.

The section 2 describes specific techniques for thickness, homogeneity and purity characterization applied to active targets during their fabrication process. The section 3 reports on methods performed during the irradiation of the material in order to monitor their thickness.

## 2 Characterization of actinide targets

### 2.1 Thickness and homogeneity

For radioactive targets, their activity at a certain time, $A(t)$, determines directly the number of atoms, $N(t)$, of a sample through its definition as $A(t) = \lambda N(t)$ knowing the decay constant, $\lambda$, of the radio-nuclide under consideration. The obtained areal density given in atoms/cm² is then straightforward applicable in the cross-sections equation.

*2.1.1 Low geometry α-particle counting (LGA)*

---

[*] Corresponding author: stodel@ganil.fr

Counting at a defined solid angle is the most accurate method for activity measurements of alpha emitters, and is also suited for x-ray emitters. The method consists of measuring with a silicon detector, the α-particle recoiling out of the active sample. Namely, the counting particles move undisturbed along a straight line from the source to the detector, other particles, i.e. those out of the opening solid angle or scattered ones are absorbed in the diaphragm placed just before the detector or in side materials (e.g. threads or baffles). All geometric parameters (distance source-detector, knife-edge aperture of the diaphragm, sample size) define precisely the solid angle and the silicon background is as low as possible. With LGA method the uncertainties as low as 0.02 % are feasible [7, 8]. As reported in [9], two different set-ups can be used according to the activity of the samples, either with a variable or a fixed source-detector distance (up to 20 cm or 1.6 m under glove box respectively) where uncertainties of 0.3-0.5 % are reached with tens of spectra acquired in about 40 hours. The distribution of the activity over the sample area is achieved by introducing an additional diaphragm other the source and moving it [10].

*2.1.2 Radiographic Imaging (RI)*

For large area targets (10-30 cm² area), Radiographic Imaging is well suited for activity measurements. Indeed, the active target is exposed in front of a phosphor screen called imaging plate (IP) which traps and stores the radiation energy. Then the IP is scanned with a laser beam which releases the energy as luminescence. This "photo-stimulated luminescence" (PSL) phenomenon is thus proportional to the amount of radiation exposed enabling to investigate the homogeneity of the target layer. The resulting spatial resolutions depend on the radioactive decay properties of the material: less than 200 μm for $^{198}$Au ($E_{β–}$ = 961 keV, $E_γ$ = 412 keV) [11, 12], 42 μm was achieved for $^{238}$U ($E_α$ = 4151 keV, $E_γ$ = 50 keV) [13]. In order to determine the absolute target activity a calibration with a reference sample is needed [12] and a thickness standard deviation of ±5.2% is obtained. This process requires about 6 hours per target, it can be applied before and after irradiation as shown in [14]. As mentioned in [15], when test experiments are performed with homologous of the actinide elements, radiotracers are used enabling the application of RI for layer homogeneity measurements.

**2.2 Isotopic Abundances**

*2.2.1 Thermal Ionization Mass Spectroscopy (TIMS)*

Thermal Ionization Mass Spectrometry (TIMS) is a highly sensitive technique to measure isotope ratio of nuclear material. It combines the thermal ionization effect of the sample to be analysed with a separation corresponding to the mass-to-charge-ratio of each isotope as performed with an electromagnet [16]. This method is applied to an aliquot of the active solution used for the electro-deposition. The atomic abundances of the solution are precisely measured with a dynamic range of the ratios up to $10^6$. The activity fractions of each isotope are then deduced and when applied to the total activity measured with LGA technique, the areal density of each active element in the target is deduced [9, 10].

*2.2.2 Neutron Activation Analysis (NAA)*

Neutron Activation Analysis (NAA) is an analytical technique based on gamma spectroscopy measurements of samples irradiated with neutron in nuclear reactors. This method is based on the identification by measuring the gamma emission of the reaction products $^{N+1}_{Z}A$ of a neutron-induced reaction $^{N}_{Z}A(n,\gamma)^{N+1}_{Z}A$, with $^{N}_{Z}A$ the isotope present in the supernatant electrolyte solution to be identified. In order to be unbiased by the total neutron flux, a reference sample can be co-irradiated under similar conditions to compare the activities of the unknown sample ($A_x$) and a known standard ($A_{st}$), the element mass ($m$) is then obtained by : $m_x = m_{st}(A_x/A_{st})$ [17]. The significant advantage of NAA is simultaneous determination of up to 30-40 elements with remarkably low detection limits, especially particular lanthanides and actinides can be evaluated in the range of 0.1-10 ppb due to their high cross-section for thermal neutron capture [18]. The process can be considered as time consuming (6 – 24h) compared to other analytical techniques [12] but fits well in cases when a direct radiometric measurement of the target is not possible since (i) the activity is far too high, (ii) in case of α-particle measurements the layer is too thick, or (iii) lateral dimensions of the target do not allow the introduction of the sample into a counter. The NAA technique is applied at Mainz to qualify actinide targets. Fig. 1 displays the γ-ray spectrum of 1 ml of $^{244}$Pu supernatant solution (extracted in the deposition cell after deposition) irradiated for 2 hours with a thermal neutron beam from TRIGA Mainz reactor at a flux of $10^{11}$ cm$^{-2}$ s$^{-1}$. After a cooling time of 0.5 hour, the γ-ray measurements are performed for 10 minutes, main γ-rays correspond to $^{245}$Am and $^{245}$Pu.

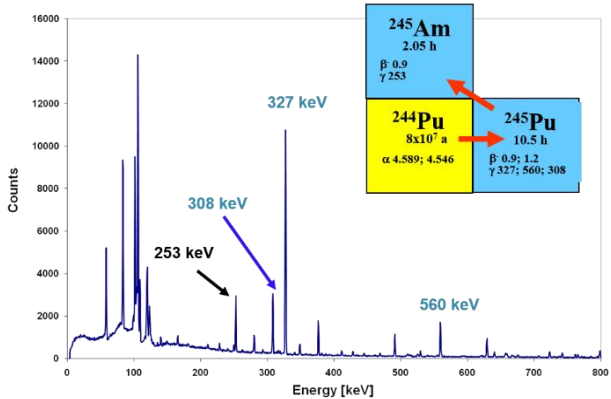

*Fig. 1. γ-ray spectrum of a $^{244}$Pu solution irradiated at the research reactor TRIGA Mainz [19].*

## 2.3 Chemical identification

### 2.3.1 Rutherford Back Scattering (RBS)

The thickness of the deposited material and the backing as well as the composition and concentration of the materials can be assigned with the Rutherford Back Scattering (RBS) method [20, 21]. A beam of particles or light ions with energy below the Coulomb barrier is impinging on targets, then scattered projectiles are registered by detectors placed at various angles to the incident beam. In order to estimate the homogeneity of the target, the measurement can be performed at different position on the surface. As shown in figure 3 of [22], the spectrum reflects the homogeneity of the deposit (long tail of the uranium low energy peak) and the presence of oxygen.

### 2.3.2 Energy Dispersive X-Ray Spectroscopy (EDS)

The chemical composition of deposited layer is obtained with energy dispersive X-ray spectrometry (EDS) using an electron beam and X-ray detectors measuring the LM-transition line of present elements [12]. This method is often coupled with a Scanning Electron Microscope in order to investigate the morphology of the targets [13].

### 2.3.3 X-Ray Photoelectron spectroscopy (XPS)

More detailed analysis of the composition and binding of the deposition is performed with X-ray photoelectron spectroscopy (XPS). It consists of scanning the target's surface with micro-focused mono-energetics X-ray source and of analysing the emitted electrons. From the shape, position and intensity of photoelectric peaks, one can deduce the chemical state and quantify the element [15, 13]. Impurities in the range of percent are analysed with the EDS and XPS methods.

## 3 Monitoring of targets under irradiation

### 3.1 Electron attenuation

By comparing the emitted current of an electron beam of 20 or 30 kV with the scattered one passing through a material, a relative thickness measurement of targets is obtained [23, 24]. When targets are mounting on a wheel, the relative thickness information over the target area is achieved by deflecting the electron beam in the radial direction (with a magnetic deflector) across the rotation, as depicted in Fig. 2.

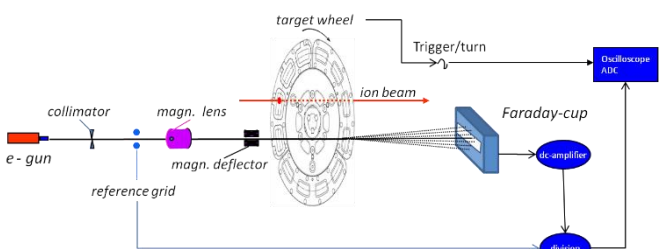

*Fig.2. Scheme of the electron gun with associated electronics and target wheel [25].*

Due to the narrow width of the electron beam, the position resolution is about 0.5 mm and the accuracy in the determination of the relative thickness is ±2%. As the electron beam, delivered by an electron gun, is parallel to the heavy ion beam used for the nuclear reaction, the targets can be analysed during irradiation, and the thickness mapping lasts at least 100 times the period of the wheel in order to achieve a spatial accuracy of 0.5 mm for targets' height of 40 mm. This method enables first to clearly identify the position of the beam on the target and secondly to evaluate the process of pin-holes or surface cracks formation according to heavy ion beam dose. With this information, the heavy ion beam position and time structure on the targets can be corrected on-line, moreover it helps in the decision of replacing the targets due to material losses [25].

## 3.2 Monitoring with α–particle emission

### 3.2.1 Energy loss

For targets thin enough to slow down alpha particles emitted by a source, the measurement of the alpha energy loss is well suited to deduce the effective thickness of the target whose composition and stopping power are known [5], [26-28]. This technique can be implemented in vacuum reaction chambers of rotating targets. By placing the wheel between the source and the detector away from the heavy ion beam direction, an on-line evaluation (depending on the source activity) of the targets thickness as a function of their length is achievable with accuracies of ±5%.

### 3.2.2 α activity

For actinide targets, when the beam is stopped and the spectrometer is tuned to the α emission velocity, the α activity of targets is measured at the focal plane. As described in [29, 30], the α spectra measured before and time to time during the experiment show various energy distributions according to the heavy ion beam dose. The position and width of the peaks reveal the changes in the targets' structure at each step of the conditioning process.

Indeed, the narrowing of the low energy tail (see figure 3 of [29]) indicates that volatile contaminants from the targets, originated from rests of solvents, resulting from the production process involving electrochemical deposition, evaporates. The saturation of the shape of the spectrum reveals that the targets' structure changes into a homogeneous transparent glass-like layer during irradiation.

As plotted in figure 6 of [30], a gradually growth of the low-energy part suggested that the target was covered by a layer of some material which was confirmed by its visual inspection.

In addition, the losses of material due to irradiation can be checked by two methods: firstly with an α source moved into the beam position in front of the target, where the counting rate of the α particles from the target is compared to the one from source; or secondly by analysing the charge equilibration foil which collects the scattered target material [29].

The α-spectra obtained at the focal plane during experiment reveals relatively large shift and broadening of the α lines compared to the ones measuring the scattered target material collected in the charge equilibration foil after experiment. The biggest contributions (estimated to be ≈60%) to these differences are due to the energy loss of the α particles in the materials (targets and detectors before the focal plane detector), nevertheless a further contribution arises to the mechanical change of the targets which become dented during irradiation [29].

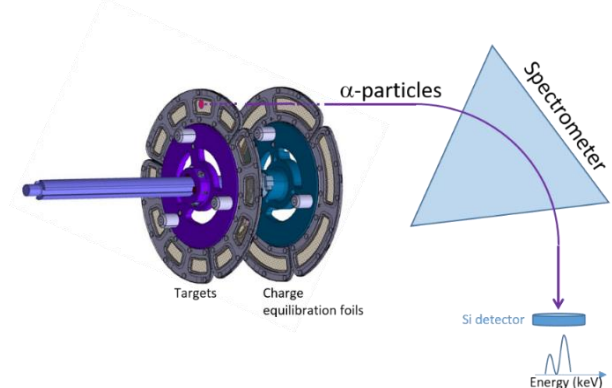

**Fig. 3.** *Scheme of the α-activity method for active targets.*

## 3.3 Rutherford scattering

One common technique to monitor the target thickness at the beam impact during irradiation is to register the elastically scattered projectiles and normalize their count rate to the beam current. The detection is performed with scintillation detectors mounted at opposite angles (± 30° in fusion-evaporation reaction) relative to the beam axis. The accuracy of the measurements depends on the precise knowledge of the solid angle of the detector.

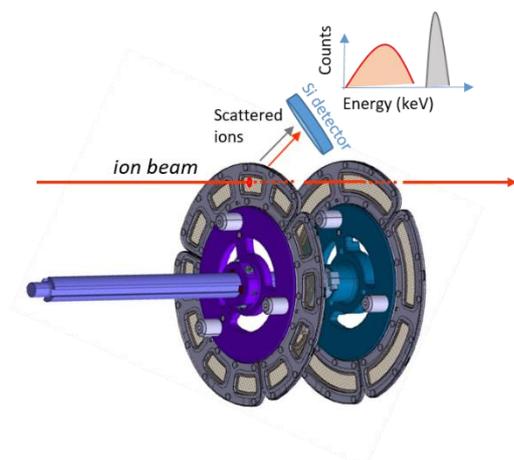

*Fig. 4. Principle of the Rutherford scattering method to monitor targets' thickness.*

### 3.4 Temperature

Due to the increase of beam intensities, targets' heating is a major issue. The target temperature can be monitored on-line by means of pyrometer [31] or infra-red devices (camera, thermo-viewer or fiberscope) [32-34]. As the targets are in a high-radiation environment, the thermal systems have to be shielded and often deported on large distance, which requires precise adjustment of the measuring position and efficient optical mirror and lenses. With the system of [33], a position resolution on targets of 0.5 mm is achieved. An essential parameter in temperature measurements is the overall emissivity on the target surface. It can be determined with heated calibrators and/or part of the surface to be analysed with known emissivity, e.g. heat-resistant black paint covering the material [34].
Complementary devices, such as CDD camera or endoscopes are used to monitor the beam spot position [33].

### 4 Summary

Due to the increasing demand for more accurate data and to the upgrade of accelerators, high quality and stable targets with precisely known properties are requested. As listed above, techniques are on common use to characterize actinide targets in order to improve the manufacture processes and to monitor and control them during irradiation in order to maintain their integrity. These techniques are also essential for the development and evaluation of new target production techniques.

*Acknowledgments.*

For their help in achieving this review and for giving me materials for the oral presentation, I want to express my gratitude to my colleagues Goedele Sibbens, Klaus Eberhardt, Bettina Lommel and Anna Solarz from European commission, Joint Research Center, Geel; Universität Mainz, Institut für kernchemie; GSI Helmholtzcentrum für Schwerionenforshung, Darmstadt and Heavy Ion Laboratory, University Warsaw respectively.